\documentclass[conference]{IEEEtran}

\usepackage{comment}
\usepackage{multirow}

\usepackage{caption}
\usepackage[subrefformat=parens]{subcaption}
\usepackage{multirow}
\usepackage{algorithm}
\usepackage{mathtools}
\usepackage{amssymb}
\usepackage{listings}
\usepackage{url}
\usepackage[noend]{algpseudocode}
\usepackage{gensymb}
\usepackage{comment}

\usepackage{cite}

\ifCLASSINFOpdf
   \usepackage[pdftex]{graphicx}
   \DeclareGraphicsExtensions{.pdf,.jpeg,.png}
\else
\fi
\begin{document}

\title{VISAS - Detecting GPS Spoofing Attacks Against Drones by Analyzing Camera's Video Stream}

\author{\IEEEauthorblockN{Barak Davidovich}
\IEEEauthorblockA{Ben-Gurion University of the Negev\\
davbarak@post.bgu.ac.il}
\and
\IEEEauthorblockN{Ben Nassi}
\IEEEauthorblockA{Ben-Gurion University of the Negev\\
nassib@post.bgu.ac.il}
\and
\IEEEauthorblockN{Yuval Elovici}
\IEEEauthorblockA{Ben-Gurion University of the Negev\\
elovici@bgu.ac.il}}

\maketitle

\begin{abstract}
In this study, we propose an innovative method for the real-time detection of GPS spoofing attacks targeting drones, based on the video stream captured by a drone's camera. The proposed method collects frames from the video stream and their location (GPS); by calculating the correlation between each frame, our method can identify an attack on a drone. We first analyze the performance of the suggested method in a controlled environment by conducting experiments on a flight simulator that we developed. Then, we analyze its performance in the real world using a DJI drone. Our method can provide different levels of security against GPS spoofing attacks, depending on the detection interval required; for example, it can provide a high level of security to a drone flying at an altitude of 50-100 meters over an urban area at an average speed of 4 km/h in conditions of low ambient light; in this scenario, the method can provide a level of security that detects any GPS spoofing attack in which the spoofed location is a distance of 1-4 meters (an average of 2.5 meters) from the real location.
\end{abstract}

\section{Introduction}
The civilian use of drones has increased in recent years, and this use is expected to grow in the future. Although drones are used in many different fields, they are vulnerable to a variety of attacks, including Global Positioning System (GPS) spoofing attacks \cite{nassi2019sok}. GPS spoofing can result in significant harm. For example, in building and construction settings, spoofing can result in the collection of incorrect locations and measurements, causing problems in the building process; in agriculture where drones are used to detect the presence of bacteria, spoofing may result in treating (spraying) the incorrect location, potentially harming crops and enabling bacteria to go untreated; and in the context of drone delivery systems, spoofing can result in delivery errors where goods are delivered to the wrong location and entities. 
Recent studies have suggested countermeasures to detect, mitigate, and prevent GPS spoofing attacks, however they have numerous disadvantages, and as a result, a recent SoK identified GPS spoofing attacks against drones as a scientific gap \cite{nassi2019sok} that threatens drones' ability to perform their tasks.

In this research, we examine whether a drone's video stream can be used to identify GPS spoofing attacks, without the need for additional drone hardware or memory, or prior knowledge of the flight area. 
We propose a method capable of detecting GPS spoofing attacks by verifying the measurements obtained by the GPS sensor against the video stream captured by a drone's camera. We analyze the suggested method's performance in a controlled environment by conducting experiments on a flight simulator that we developed. Then, we analyze its performance in the real world using a commercial drone. We show that our method can identify any attempt to launch GPS spoofing attack in which the spoofed location is a distance of 1-4 meters (an average of 2.5 meters) from the real location, for a drone flying at an altitude of 50-100 meters over an urban area at an average speed of 4 km/h with different levels of ambient light.

The significance of our method with respect to the methods proposed in related work is as follows. Our method (1) relies on existing hardware: in contrast to methods presented in other studies (e.g., \cite{abera2019diat}, \cite{jansen2016multi}, and \cite{ranganathan2016spree}), our method does not involve the use of additional hardware which makes the method also cost-effective; (2) is database independent: in contrast to methods presented in other studies (e.g., \cite{xue2020deepsim}) which use a precompiled database, our method does not rely on a precompiled database or a map of the drone's flight area;  and (3) our method offers flexibility: unlike other methods, it can be implemented either on the drone or at the ground control station used to control the drone (i.e., on the drone's controller). In addition (4), we empirically evaluate the accuracy of our method and determine the level of security for a situation in which the spoofed location is an average of 2.5 meters away from the actual location, an aspect that was not evaluated in related studies.

The remainder of this paper is organized as follows. 
We provide an overview of related work in Section \ref{sec:related-work}. 
The proposed method is described in Section \ref{sec:Proposed Method}. In Section \ref{sec:analysis}, we discuss the analysis performed in our simulation environment, and the results of our real-world evaluation are presented in Section \ref{sec:evaluation}. Our paper concludes in Section \ref{sec:summary}, where we summarize and discuss our findings.

\section{Related Work}
\label{sec:related-work}

In this section, we review prior studies that proposed countermeasures against GPS spoofing. 
The GPS protocol is vulnerable to spoofing attacks, since it lacks encryption and authentication mechanisms. 
As a result, attackers can inject false GPS signals using software-defined radio (SDR) or dedicated GPS spoofers (which can be bought online), causing the drone to believe that it is flying in a location that differs from its actual location.
Various studies have shown that GPS spoofing attacks against drones can cause a drone flying in autonomous mode to accelerate in the desired direction chosen by the attacker \cite{kerns2014unmanned} by transmitting a fake GPS coordinates in the opposite direction which causes acceleration in the original flight direction, or force a drone flying in manual mode to land \cite{luo2016drones} (by sending a no-flight zone alert that triggers a safety mechanism, causing the drone to land).

Various studies have suggested countermeasures against GPS spoofing attacks by integrating additional hardware (e.g., \cite{jansen2016multi}, \cite{ranganathan2016spree}). 
One study \cite{jansen2016multi} suggested the use of a multi-receiver for GPS spoofing detection to detect malicious fake signals, by verifying the GPS measurements using the fixed distances between the receivers and then measuring the distances between the receivers’ reported locations. When the GPS signal is legitimate, the distance will be similar to the fixed distances, but when there is GPS spoofing attack, the measured distances will be very close to zero, as all the receivers are spoofed with the same fake location; this method would be difficult to implement with small drones, because additional hardware would be needed for all of the GPS receivers. SPREE \cite{ranganathan2016spree}, a method presented in another study, is a countermeasure for GPS spoofing attacks that can also detect takeover attacks; it relies on the auxiliary peak method used in combination with a navigation message inspector, in which the strongest satellite signal as well as other weaker environment signals are tracked. SPREE's main disadvantage is that external hardware is needed.

Other studies suggested countermeasures that use existing hardware (e.g., \cite{abera2019diat}, \cite{feng2018efficient}, \cite{xue2020deepsim}, and \cite{qiao2017vision}). Several proposed methods use motion sensors and compasses to detect GPS spoofing attacks. For example, one study \cite{feng2018efficient} presented a method that uses gyroscope measurements to verify GPS measurements; its drawbacks are that the sensor's measurements suffer from false negative/positive errors and must be calibrated in advance. Another method \cite{qiao2017vision} uses an on-board camera and the inertial measurement unit (IMU) to obtain the velocity and position of the drone to detect unexpected changes in the flight path. In this case, the drawbacks are the same as the previous method; additionally, this method relies on two sensors: the camera and IMU. A deep learning-based solution which uses images from satellites and compares them to images from a drone's camera to determine whether the locations match was proposed by \cite{xue2020deepsim}; its disadvantage is that it requires initial preparation in the flight area. Moreover, every GPS point in the area must be covered, a requirement which increases the size of the precompiled database significantly. Another study presented DIAT\cite{abera2019diat}, which verifies the data/signals received from a drone's sensors with data from other nearby drones in order to detect compromised measurements. Its drawback is that it needs to support and protect drone-to-drone communication, which requires development of the communication protocol. 

While various countermeasures against GPS spoofing attacks have been suggested, they all have disadvantages. A recent SoK paper \cite{nassi2019sok} identified GPS spoofing attacks against civilian drones as a scientific gap that cannot be prevented by any mechanism with a high technological readiness level \cite{wiki:Technology_readiness_level}.

\section{Proposed Method}
\label{sec:Proposed Method}

In this section, we describe the proposed method for the detection of GPS spoofing attacks on drones. Our method relies on data obtained from two sources: a drone's video stream and GPS measurements. It is based on the assumption that unlike GPS signals, video streaming from the drone cannot be spoofed. To detect GPS spoofing attacks, our method correlates a drone's movement, calculated from the GPS signals, with the real-time video stream frames. Based on this correlation and a predefined  correlation threshold, our method determines whether a GPS spoofing attack occurred.

The difference in the correlation between the first frame ($frame_{i}$) and the next n consecutive frames ($frame_{i+1}$, ......,$frame_{i+n}$) is continuously calculated.
Based on the GPS measurements and the similarity correlation between the frames, a model, which is used to verify the location for the next q consecutive frames ($frame_{i+n+1}$, ......,$frame_{i+n+q}$), is created.
For each of the next q consecutive frames ($frame_{i+n+1}$, ......,$frame_{i+n+q}$), the distance, is predicted, based on the frame's similarity correlation with the first frame. 
If the error, meaning the difference between the actual distance and the distance predicted by the model, is beyond a threshold, the model issues an alert that the GPS measurements do not correlate with the video stream (i.e., a GPS spoofing attack has taken place). 
In each flight, our method is applied for a specified period of time (between $frame_{i}$ to $frame_{i+n+q}$). The similarity correlation between consecutive frames decreases as a function of the distance, so eventually there will be no correlation between the first and last frames, and a new model will be generated.

\begin{figure}[ht]
\begin{center}
    \fbox{\includegraphics[width=.35\textwidth]{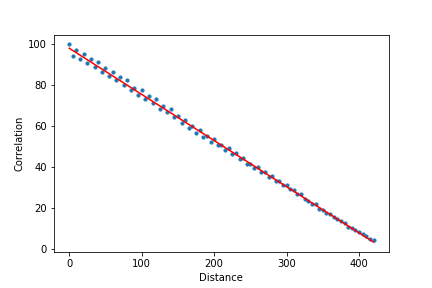}} \hspace{1PX}
    \caption{Correlation as a function of distance for location \#1 (a neighborhood in Manhattan) and an altitude of 200 meters (the blue points indicate the simulation results, and the red line represents the linear regression).} \label{fig:example 200m}
\end{center}
\end{figure}

\begin{figure*}[ht]
\centering
    \includegraphics[width=1\textwidth]{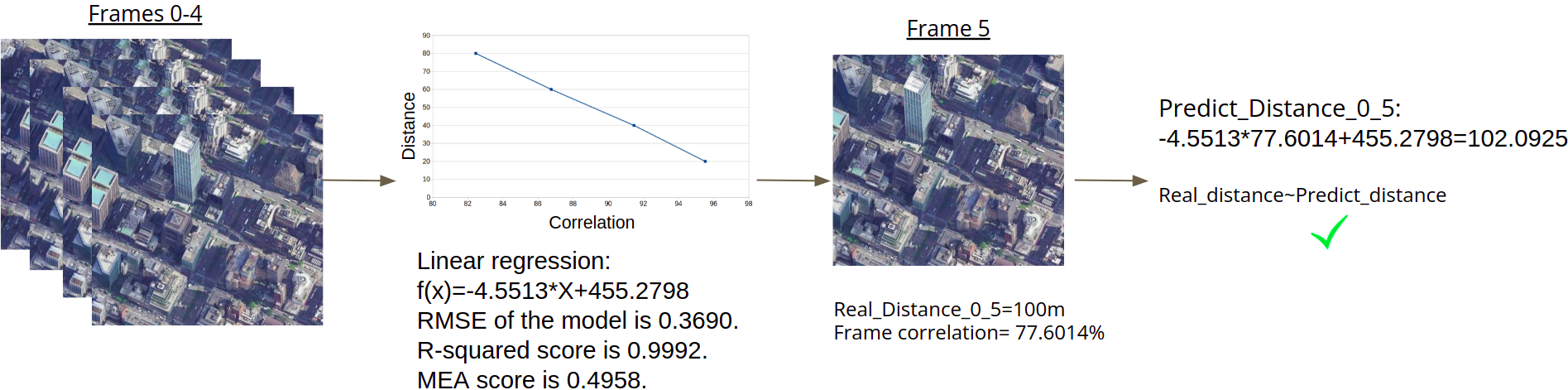}    
    \caption{The implementation of the proposed method, using an example in which we get 5 frames (frame\_0 to frame\_4) and try to predict the distance for frame\_5 (i=0, n=4).}
    \label{fig:the_metod}
\vspace{-1.5em}
\end{figure*}

Our method includes the following steps: (1) After the drone takes off, frames along with their specific GPS locations, are collected. For each second of flight, one frame and its GPS location are saved. Thus, after n seconds, we will have n frames (frame\_i = the first frame in the time window, and frame\_i+n = the $n^{\text{\tiny th}}$ frame in the time window). (2) The similarity correlation between frame\_i and all other i+n frames is calculated. At the same time, the distance (based on the GPS measurements) between the location of frame\_i and the location of all other i+n frames is calculated. This information is used to generate a graph presenting correlation vs distance (an example of such a graph is presented in Figure \ref{fig:example 200m}). This graph allows us to model the change in frame correlation against the GPS distance, which provides a suitable function for the prediction of the next point ($f(correlation)=distance$) on the graph. (3) The next frame i+n+1 and its GPS location are obtained, and the correlation between frame\_i and frame\_i+n+1 ($correlation\_i\_i+n+1$) and the distance between the GPS locations are calculated. Then, the model can predict $predict\_distance$ from function $f(correlation\_i\_i+n+1)$ = $predict\_distance$. We then have the actual and predicted distance between frame\_i and frame\_i+n+1. (4) If $predicted\_distance \approx real\_distance$, the GPS location correlates with the frame, indicating that there was no GPS spoofing attack on the drone. If the $predicted\_distance$\textgreater$real\_distance$, there is no correlation between the GPS location and the frame, confirming that there has been a GPS spoofing attack. This process is presented in Figure \ref{fig:the_metod} using an example in which we collect five frames (frame\_0 to frame\_4), along with their GPS locations, and tried to predict the distance for frame\_5 (i=0, n=4).

\section{Analysis and Simulation}
\label{sec:analysis}

In this section, we describe the experiments and analysis performed in our simulation environment, a setting which allows us to investigate the effect of (1) the drone's altitude, (2) the drone's speed, (3) the terrain over which the drone flies, and (4) ambient light on our method's performance. The simulation environment was designed so that we could examine our solution in a controlled environment without external disturbances and demonstrate a proof of concept of our solution in various field conditions. In this setting we can also test our proposed method anywhere in the world by using Google Earth. To predict the next position, we graphed correlation vs distance and built a function suitable for the graph and learn how it behaves. A test window is defined as the number of points needed to construct a linear regression function so that $f(correlation) = distance\_prediction$.

We used the simulator to conduct various experiments in order to assess the influence of the specific factors (1-4 in the previous paragraph) on the method's performance. The following metrics were used to evaluate the performance: the root-mean-square-error (RMSE), $R^2$, and mean absolute error (MAE).

\begin{figure*}[ht] 
\begin{center}
    \includegraphics[width=.41\textwidth]{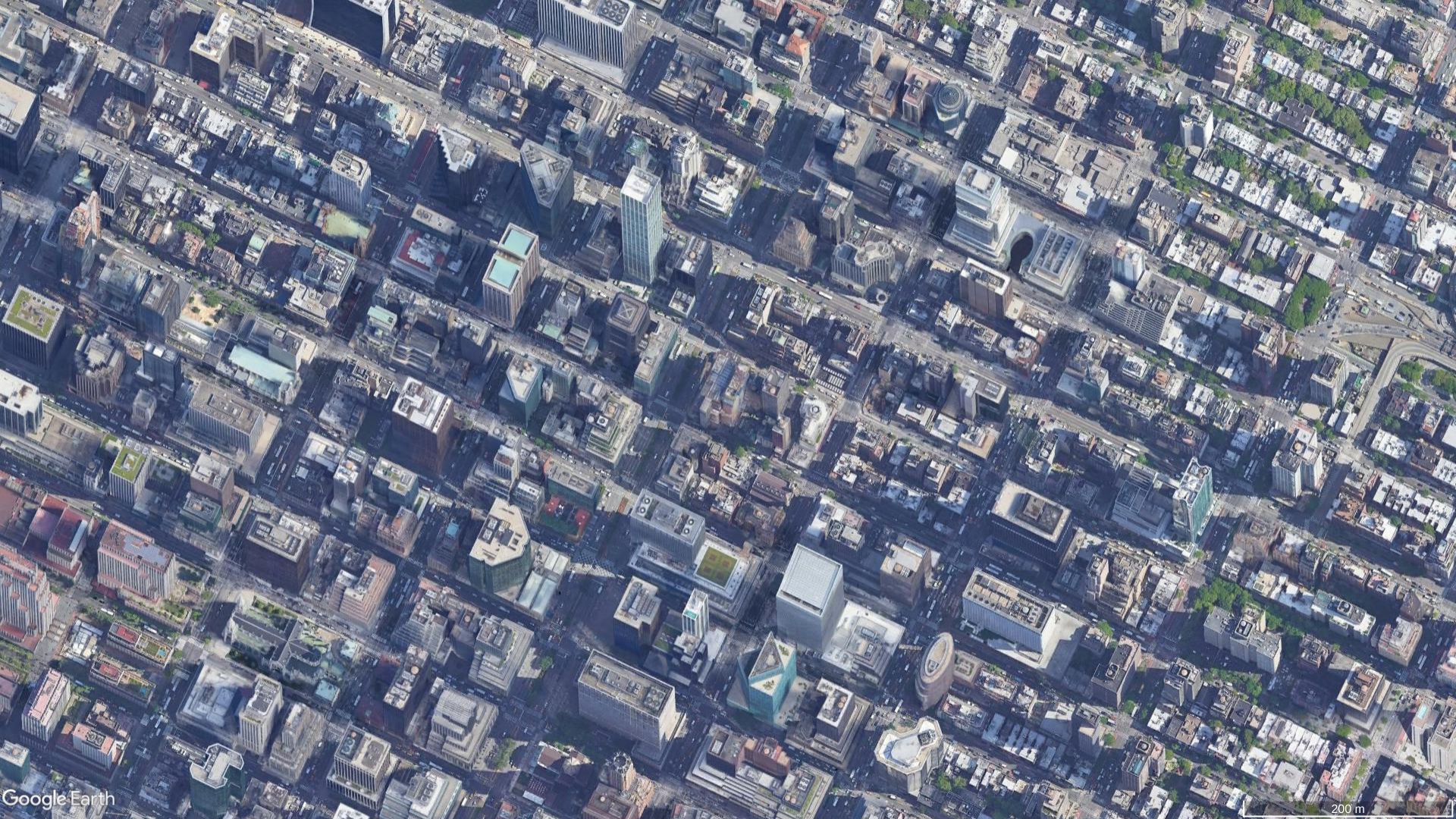}  \hspace{1PX}
    \includegraphics[width=.41\textwidth]{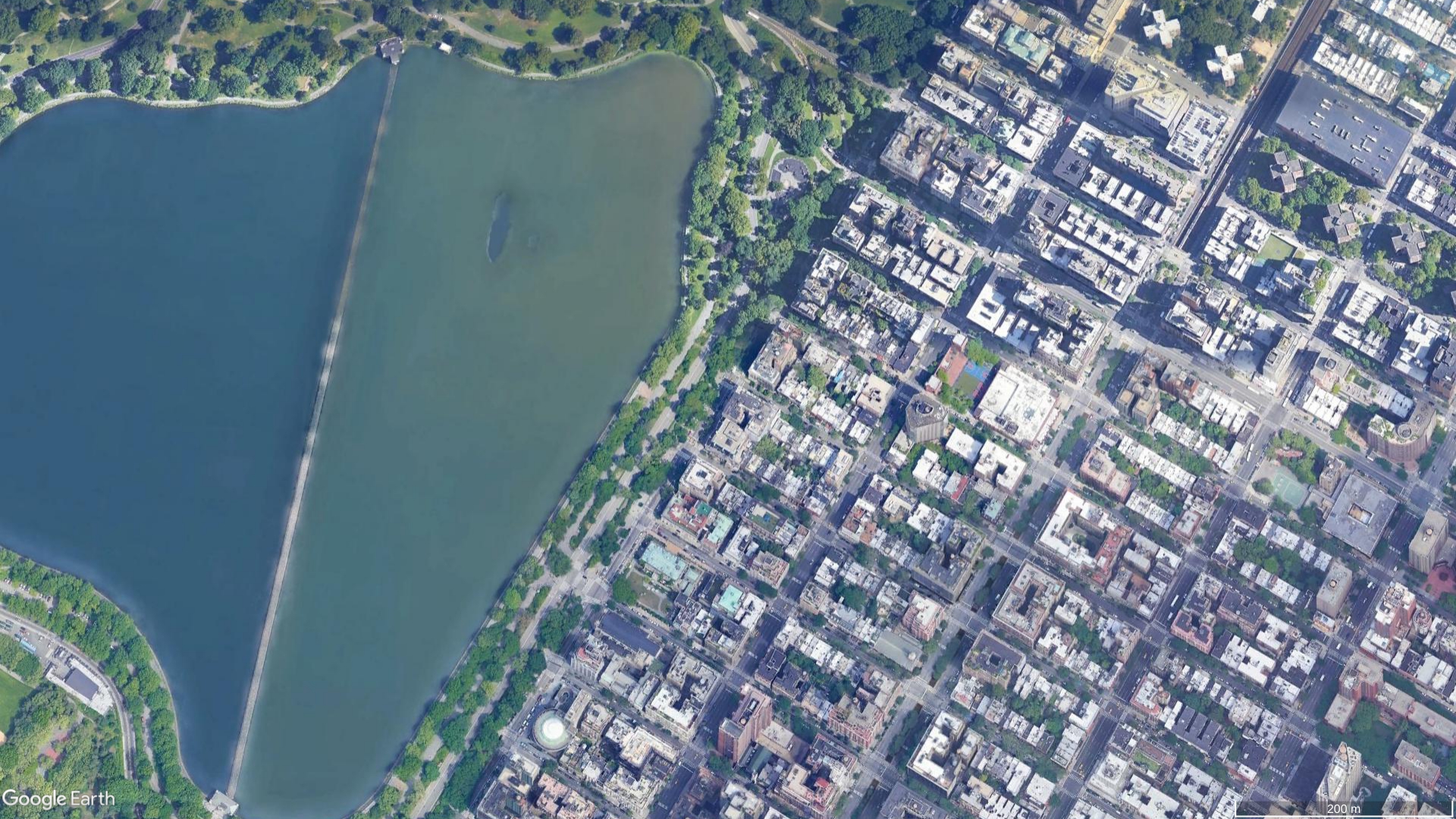}      \hspace{1PX}
    \caption{The view of the simulated environment: left - a neighborhood in Manhattan (location \#1) and right - the lake in Central Park (location \#2).} 
    \label{fig:locations simulation}
\end{center}
\end{figure*}

\subsection{Influence of altitude}
\label{sub:Influence of Height}

The altitude affects the efficiency of the test window and how quickly it needs to be changed. As the drone's altitude increases, its perspective widens, and the changes between two consecutive frames are more significant. 

Experimental Setup: In this experiment, we chose a location in a neighborhood of Manhattan (location \#1 in Figure \ref{fig:locations simulation}). In the simulations, the drone flew at three different altitudes (50, 100, and 200 meters) at a consistent speed of 5 m/s (meters per second). One frame was collected per second. For each altitude, the simulation ended immediately after the drone was out of the scope of the first frame (this was done in each of the experiments described below), because we found that a correlation of zero is obtained when the drone is out of the first image's scope due to the fact that there is no longer any similarity to the first image. 

Results: As can be seen in Table \ref{tab:Hight test location 1}, similar results were obtained for the three altitudes examined. Several interesting insights can be derived from the results: (1) RMSE: as the altitude increases, the RMSE increases; (2) $R^2$: while this tends to be a value of one, it can be seen that the $R^2$ decreases at $10^-3$; (3) MEA: when the value is close to zero we get a better match to reality.

\begin{table}[ht]
\centering
\resizebox{\columnwidth}{!}{%
    \begin{tabular}{|c|l|l|l|l|} 
    \hline
    Altitude & \multicolumn{1}{c|}{Linear function} & \multicolumn{1}{c|}{RMSE} & \multicolumn{1}{c|}{R2} & \multicolumn{1}{c|}{MEA}  \\ 
    \hline
    200M  & -4.4377*X+435.1809 & 24.8307 &  0.9984 &  4.1468 \\ 
    \hline
    100M  & -2.1408*X+199.966 & 17.5184 & 0.9952 & 3.0840    \\ 
    \hline
    50M   & -0.9627*X+96.0829 & 10.4591 & 0.9885 &  2.4496        \\
    \hline
    \end{tabular}
    }
\caption{The effect of a drone's altitude on the performance; the results are presented for location \#1 and various altitudes, with a drone speed of 5 m/s}
\label{tab:Hight test location 1}
\end{table}

\subsection{Influence of Speed}
\label{sub:Influence of Speed}
The drone's speed can affect the efficiency of the test window and how quickly it needs to be changed. As the drone's speed increases, its perspective widens, and the correlation between consecutive frames decreases accordingly. To isolate the problem, a constant velocity is used in this experiment.

Experimental Setup: In this experiment, we used one location in Manhattan (location \#1 in Figure \ref{fig:locations simulation}) and simulated a drone flight at an altitude of 200 meters. For each altitude, three different speeds and a constant velocity were used. One frame was collected per second. As mentioned in Section \ref{sub:Influence of Height}, the simulation ended immediately after the drone was out of the scope of the first image.

Results: Table \ref{tab:speed 200M} presents the results for an altitude of 200 meters. Several interesting insights can be derived from the results: (1) RMSE: as the speed increases, the RMSE decreases; (2) $R^2$: while this tends to be a value of one, it can be seen that the $R^2$ decreases at $10^-3$; (3) MEA: when the value is close to zero we get a better match to reality; and (4) by examining each speed separately, we observe that the best speed for this terrain is 10 m/s, where it can be seen that the values of the RMSE, $R^2$, and MEA are the lowest for all altitudes.

\begin{table}[ht]
    \resizebox{\columnwidth}{!}{%
    \begin{tabular}{|l|l|l|l|l|}
    \hline
    \multicolumn{1}{|c|}{Speed} & \multicolumn{1}{c|}{Linear function} & \multicolumn{1}{c|}{RMSE} & \multicolumn{1}{c|}{R2} & \multicolumn{1}{c|}{MEA} \\ \hline
    5 m/s & -0.2257*X+98.1342 & 1.2764 & 0.9983 & 0.9385 \\ \hline
    10 m/s & -0.2305*X+99.8311 & 0.2852 & 0.9996 & 0.4229 \\ \hline
    20 m/s & -0.2294*X+99.9467 & 0.5753 & 0.9993 & 0.5700 \\ \hline
\end{tabular}
}
\caption{The effect of a drone's speed on the performance; the results are presented for location \#1 and an altitude of 200 meters, with various speeds}~\label{tab:speed 200M}
\end{table}

\begin{figure*}[ht] 
  \begin{subfigure}[b]{0.5\linewidth}
    \centering
    \includegraphics[width=0.75\linewidth]{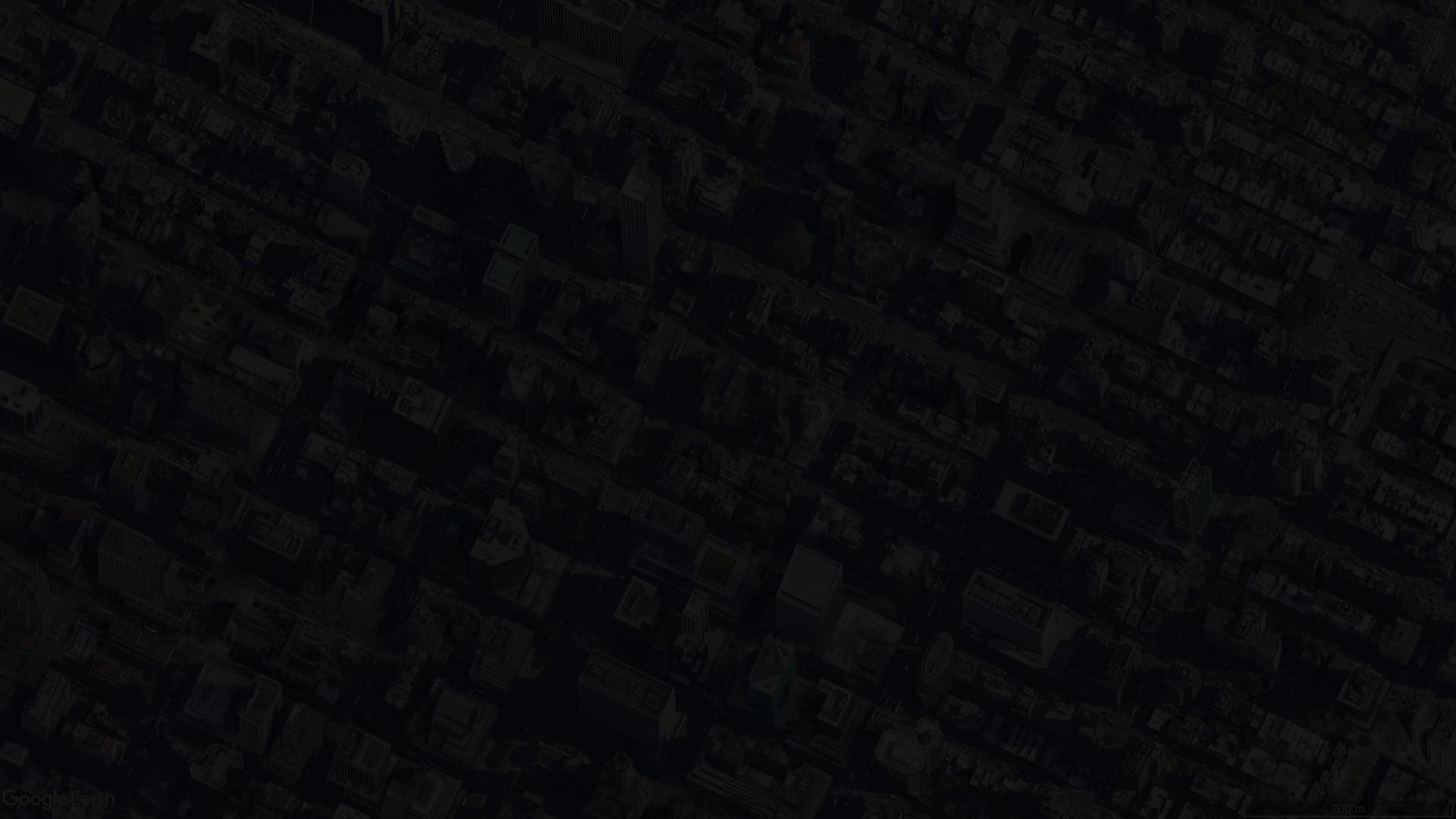} 
    \caption{10\% light} 
    \vspace{4ex}
  \end{subfigure}
  \begin{subfigure}[b]{0.5\linewidth}
    \centering
    \includegraphics[width=0.75\linewidth]{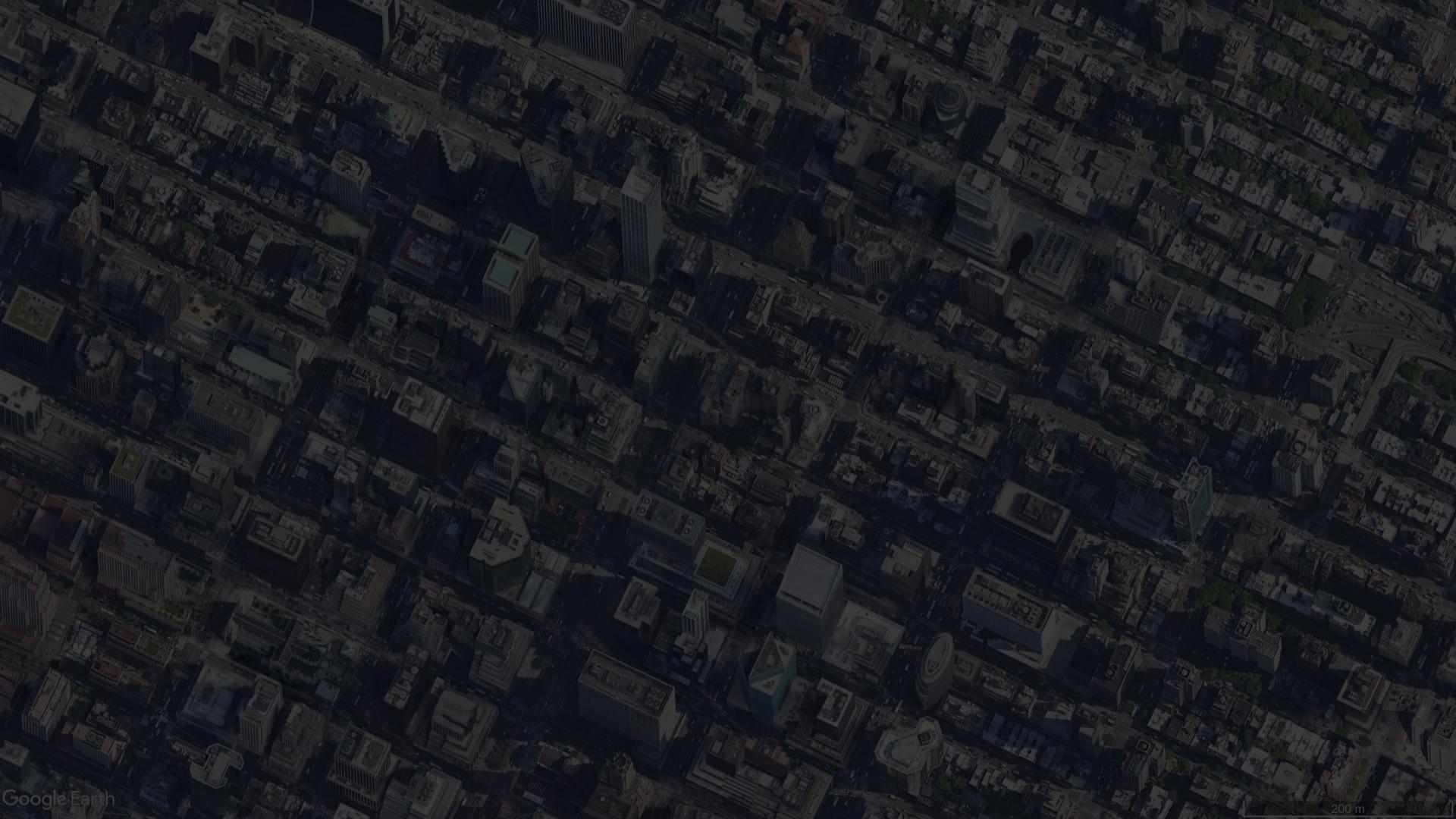} 
    \caption{25\% light} 
    \vspace{4ex}
  \end{subfigure} 
  \begin{subfigure}[b]{0.5\linewidth}
    \centering
    \includegraphics[width=0.75\linewidth]{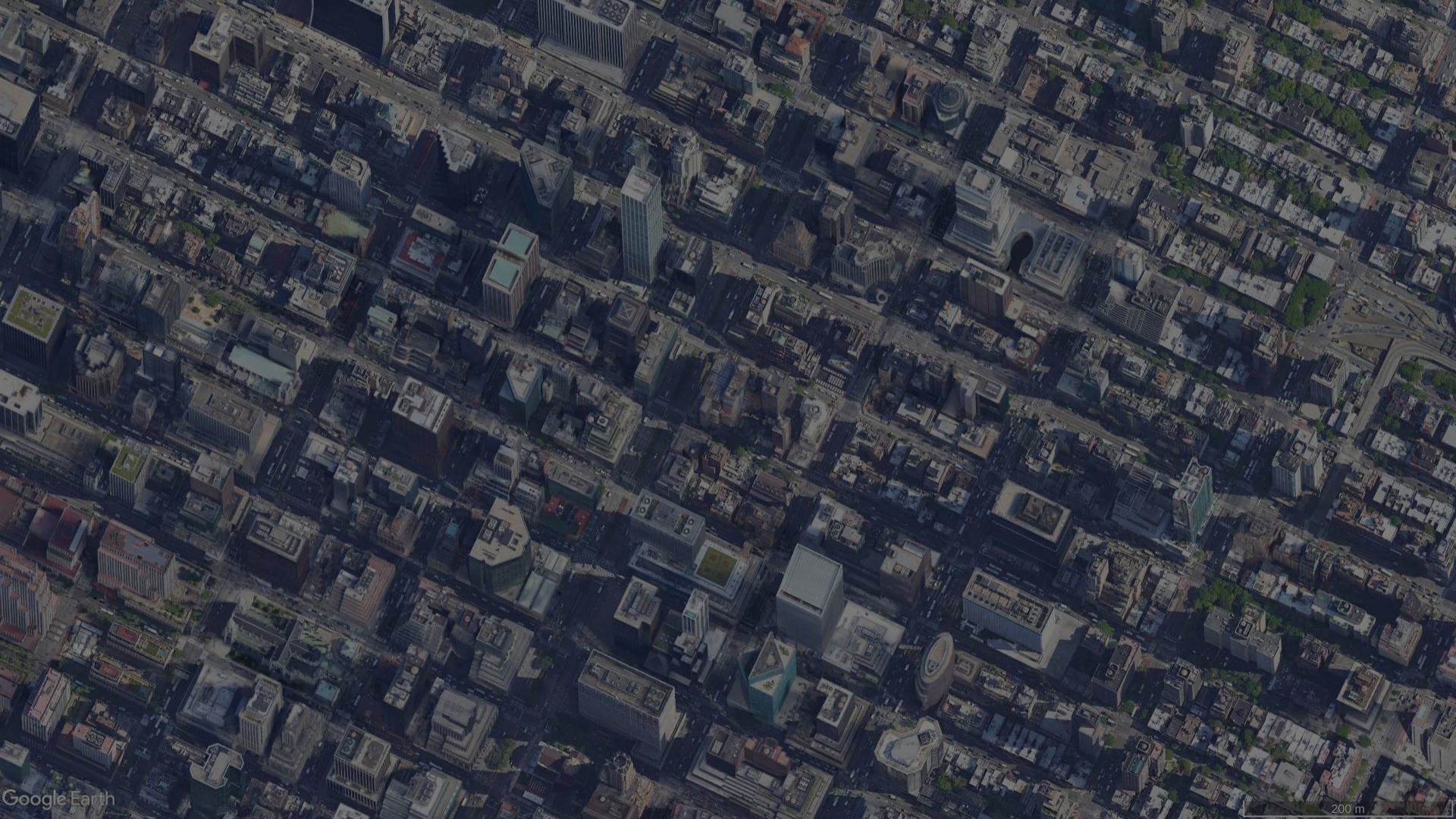} 
    \caption{50\% light} 
  \end{subfigure}
  \begin{subfigure}[b]{0.5\linewidth}
    \centering
    \includegraphics[width=0.75\linewidth]{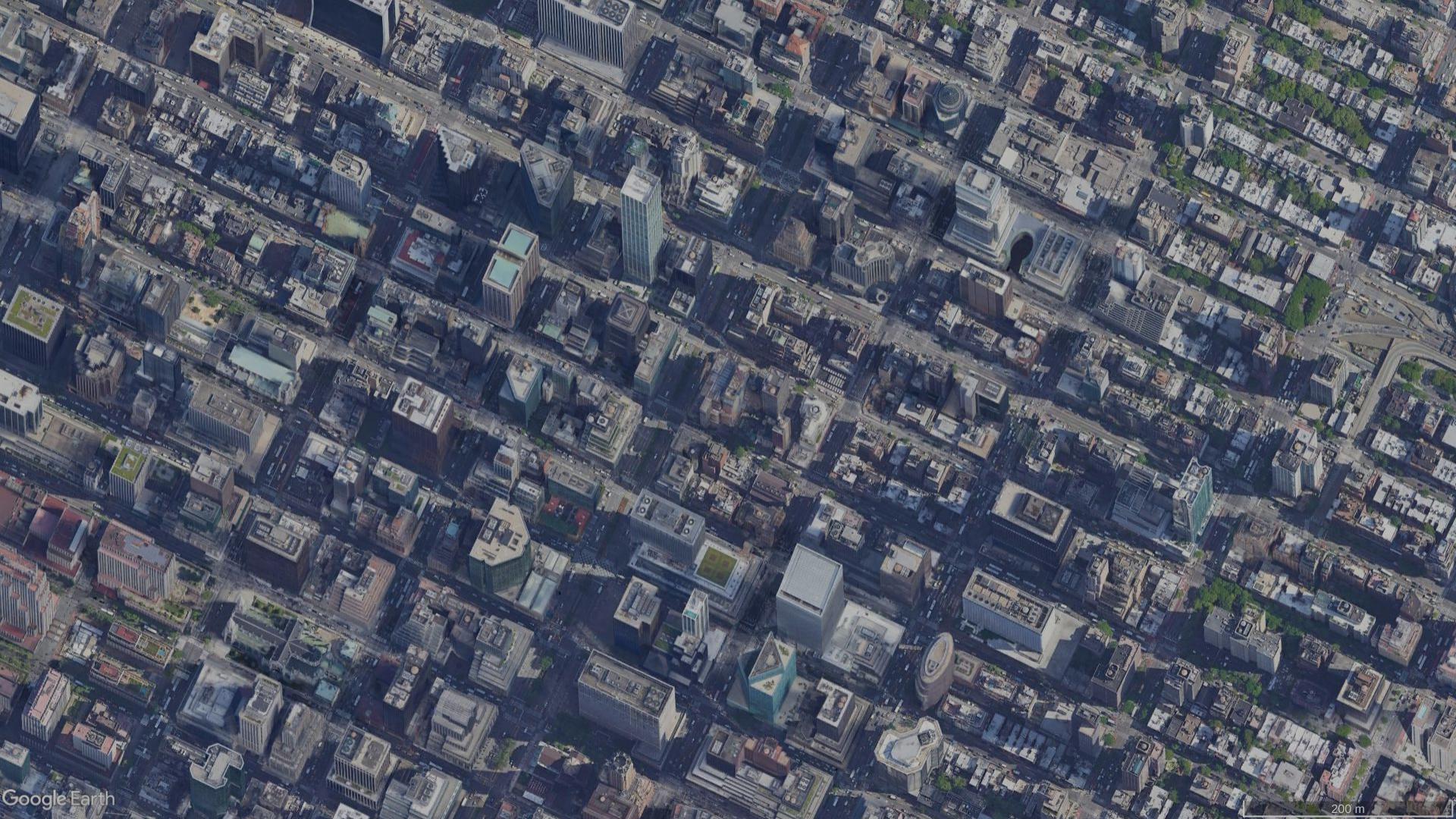} 
    \caption{75\% light} 
    
  \end{subfigure} 
  \caption{The view of an urban location (location \#1) from an altitude of 1.2 kilometers in various lighting conditions (ranging from 75 to 10\% light).}
  \label{fig:light} 
\end{figure*}

\subsection{Influence of Terrain}
\label{sub:Influence of Terrain}
When the terrain over which the drone flies changes constantly (as it does in urban areas), the correlation between the frames has greater influence; in contrast in unsettled open/flat areas there are limited changes in the correlation between consecutive frames. The results presented in Section \ref{sub:Influence of Height} demonstrate that the proposed method provides good results in urban environments, so in this section, we examine only unsettled open/flat areas. 

Experimental Setup: In this simulation, the drone flew over the lake in Central Park (location \#2 in Figure \ref{fig:locations simulation}). As the starting point, we used the worst-case scenario in which the drone hovers over the lake at three different altitudes (200, 100, and 50 meters) at speeds of 5, 10, and 20 m/s. As mentioned in Section \ref{sub:Influence of Height}, the simulation ended immediately after the drone was out of the scope of the first image

Results: Table \ref{tab:terrain 200M} presents the results for this terrain for an altitude of 200 meters (note that the same conclusion can be drawn based on the results for each of the altitudes examined).  The results obtained for the three metrics indicate that none of the metrics were suitable for evaluating the influence of the terrain on our method's performance. However, several interesting insights can be derived from the results: (1) RMSE: the RMSE value is very high (over 180) at all speeds and altitudes; (2) $R^2$: the value obtained for this metric is never close to one; therefore the error keeps increasing, and (3) MEA: the MEA value is never close to zero. Based on the results of this experiment, we conclude that our solution is not effective when applied on unsettled open/flat terrains; it is more suitable for urban terrain.

\begin{table}[ht]
    \resizebox{\columnwidth}{!}{%
    \begin{tabular}{|c|l|l|l|l|}
    \hline
    Speed & \multicolumn{1}{c|}{Linear function} & \multicolumn{1}{c|}{RMSE} & \multicolumn{1}{c|}{R2} & \multicolumn{1}{c|}{MEA} \\ \hline
    5m/s  & -0.2942*X+118.5731                   & 300.9058                  & 0.8159                  & 15.1959                  \\ \hline
    10m/s & -0.2796*X+115.8056                   & 278.9850                  & 0.8254                  & 14.5698                  \\ \hline
    20m/s & -0.2663*X+113.0726                   & 290.8093                  & 0.8237                  & 14.8485                  \\ \hline
    \end{tabular}
    }
\caption{The effect of the flight terrain on the performance; the results are presented for the Central Park lake location and an altitude of 200 meters, with various speeds}~\label{tab:terrain 200M}
\end{table}

\subsection{Influence of Ambient Light}
Ambient light can influence how the terrain trajectory changes. We note that our method relies on the video stream from a drone's built-in camera, which does not provide any special night vision capabilities or the ability to see in total darkness. 

Experimental Setup: To obtain images with different levels of daylight, which are not available from Google Earth, we used a Python technique commonly used to darken images. This allowed us to simulate four lighting conditions: 75\%, 50\%, 25\%, and 10\% light (see Figure \ref{fig:light}) at altitudes of 50-200 meters. In this experiment, one location was used (location \#1), along with the four lighting conditions. As mentioned in Section \ref{sub:Influence of Height}, the simulation ended immediately after the drone was out of the scope of the first image. 

Results: Once again, several interesting observations can be derived from the results, which are presented in Table \ref{tab:light}: (1) RMSE: we observe that when there is less light (reflected in the darkened images), the RMSE increases; (2) $R^2$: as the amount of light decreases, we observe a small decrease in the values for this metric; (3) MEA: as the amount of light decreases, the MEA value increases; (4) for all lighting changes, the best results obtained are for an altitude of 200 meters; (5) poorer results are obtained when the lighting level is under 25\%. Our solution relies on the fact that there is a change in terrain every frame transition; these changes cannot be detected in dark conditions, which is why the ambient light conditions affects our solution.

\begin{table}[ht]
\resizebox{\columnwidth}{!}{%
    \begin{tabular}{|c|c|c|c|c|c|}
    \hline
    Altitude               & Light & Linear function    & RMSE     & R2     & MEA     \\ \hline \hline
    \multirow{4}{*}{50M}  & 75\%  & -1.0636*X+99.5752  & 3.0514   & 0.9967 & 1.4503  \\ \cline{2-6} 
                         & 50\%  & -1.1111*X+102.3809 & 6.8292   & 0.9926 & 2.1720  \\ \cline{2-6} 
                         & 25\%  & -1.1123*X+101.8349 & 7.1568   & 0.9923 & 1.9989  \\ \cline{2-6} 
                         & 10\%  & -1.0263*X+125.1315 & 543.1094 & 0.5925 & 19.7922 \\ \hline \hline
    \multirow{4}{*}{100M} & 75\%  & -0.4858*X+96.4901  & 3.4121   & 0.9956 & 1.4333  \\ \cline{2-6} 
                         & 50\%  & -0.4863*X+95.7304  & 4.5149   & 0.9943 & 1.6018  \\ \cline{2-6} 
                         & 25\%  & -0.5002*X+99.1480  & 4.7360   & 0.9943 & 1.8010  \\ \cline{2-6} 
                         & 10\%  & -0.4757*X+123.3902 & 246.8611 & 0.7533 & 13.2136 \\ \hline \hline
    \multirow{4}{*}{200M} & 75\%  & -0.2261*X+98.8673  & 1.3956   & 0.9981 & 0.9657  \\ \cline{2-6} 
                         & 50\%  & -0.2310*X+99.0908  & 2.4037   & 0.9970 & 1.2434  \\ \cline{2-6} 
                         & 25\%  & -0.2351*X+100.4920 & 2.7024   & 0.9967 & 1.2696  \\ \cline{2-6} 
                         & 10\%  & -0.2258*X+103.0895 & 41.5135  & 0.9487 & 5.8316  \\ \hline
    \end{tabular}
}
\caption{The effect of ambient light on the performance; the results are presented for an urban location, an altitude of 1.2 kilometers, and various lighting conditions (ranging from 75 to 10\%). }
\label{tab:light}
\end{table}

\begin{figure*}[ht] 
  \begin{subfigure}[b]{0.5\linewidth}
    \centering
    \includegraphics[width=0.75\linewidth]{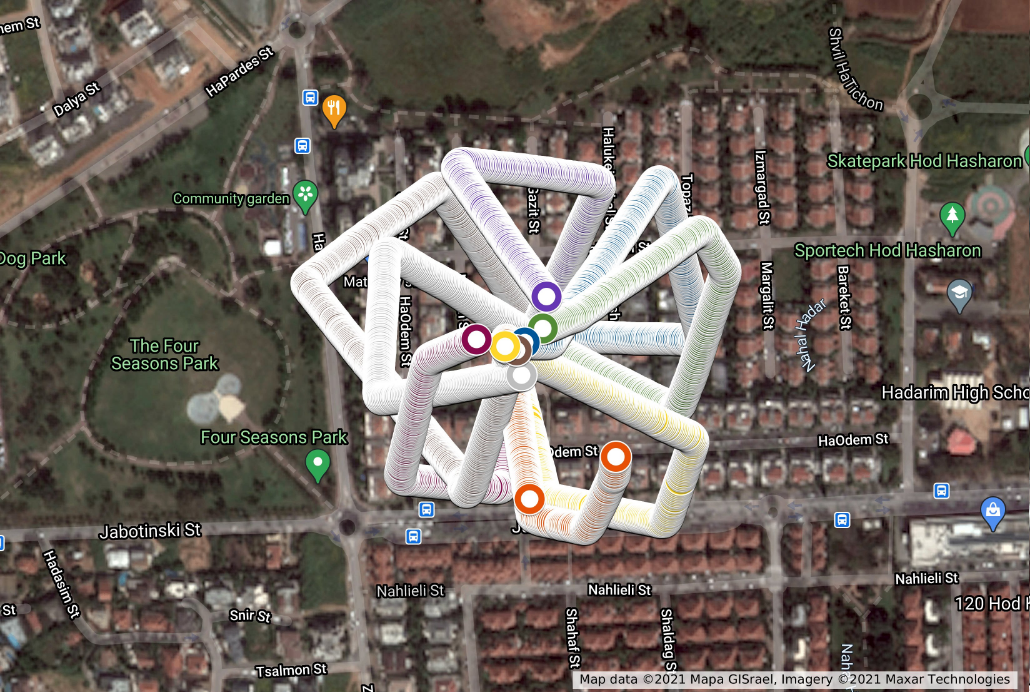} 
    \vspace{4ex}
  \end{subfigure}
  \begin{subfigure}[b]{0.5\linewidth}
    \centering
    \includegraphics[width=0.75\linewidth]{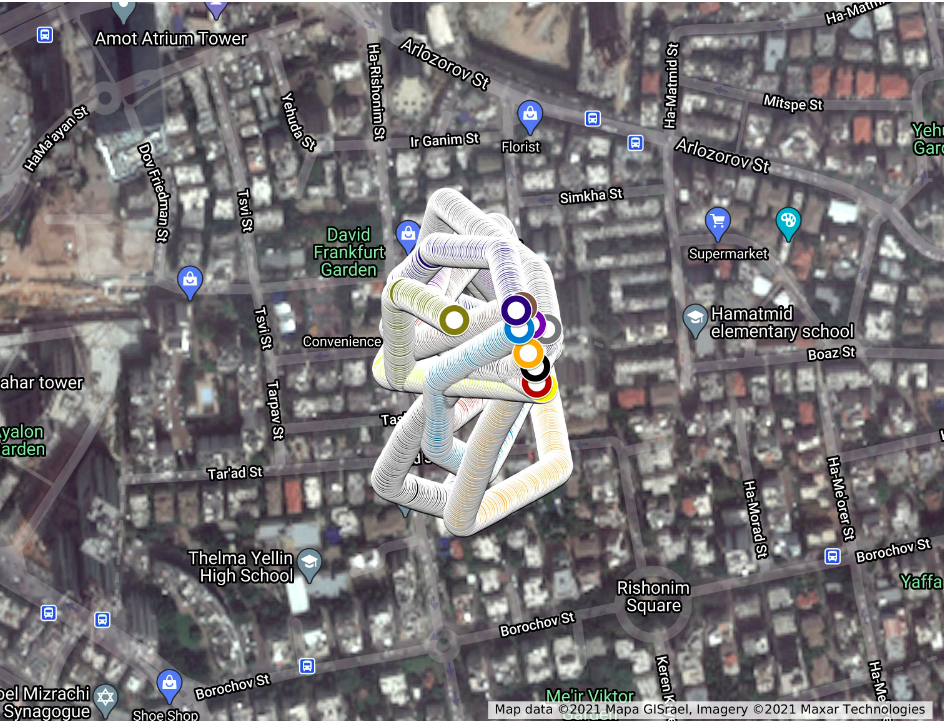} 
    \vspace{4ex}
  \end{subfigure} 
  \caption{The drone flight route for location \#1 (left) and location \#2 (right) at an altitude of 50 meters.}
  \label{fig:hod_ramat_GPS_50m} 
\end{figure*}

\section{Real-World Evaluation}
\label{sec:evaluation}

\begin{figure*}[ht] 
  \begin{subfigure}[b]{0.5\linewidth}
    \centering
    \includegraphics[width=0.75\linewidth]{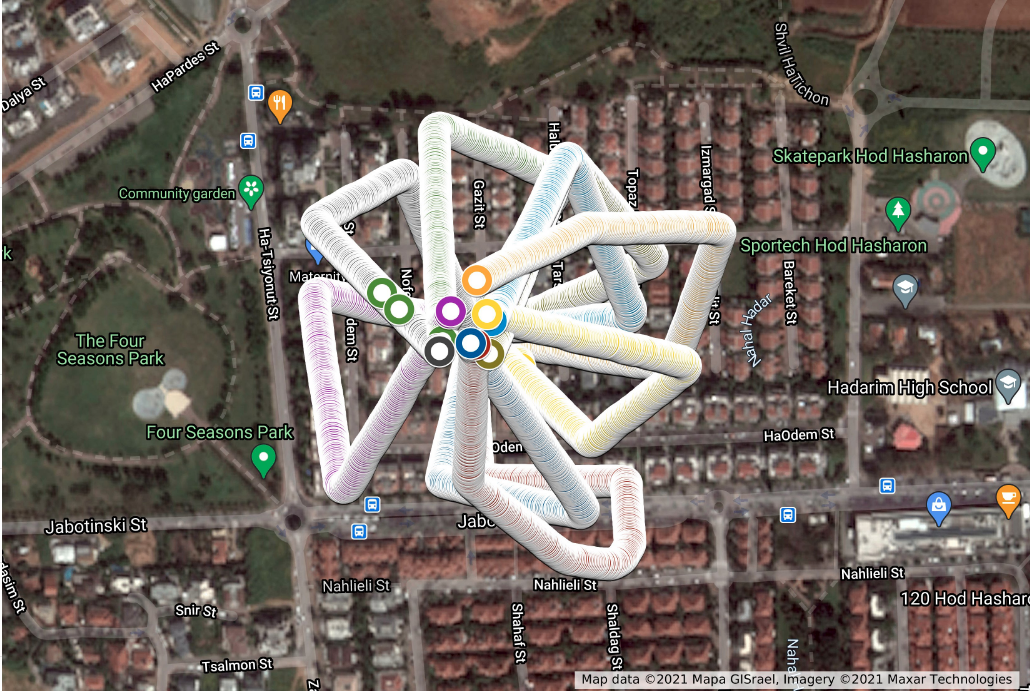} 
    \vspace{4ex}
  \end{subfigure}
  \begin{subfigure}[b]{0.5\linewidth}
    \centering
    \includegraphics[width=0.75\linewidth]{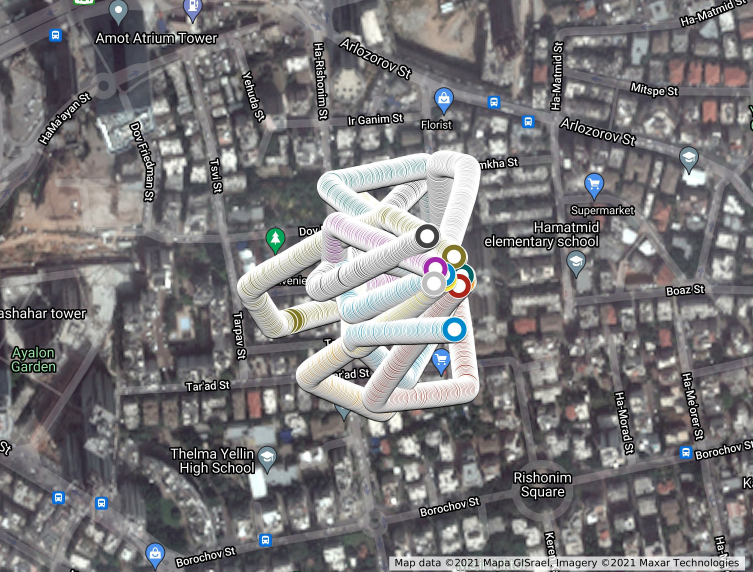} 
    \vspace{4ex}
  \end{subfigure} 

  \caption{The drone flight route for location \#1 (left) and location \#2 (right) at an altitude of 100 meters.}
  \label{fig:hod_ramat_GPS_100m} 
\end{figure*}

In this section, we describe our real-world evaluation of the proposed method using a DJI Mavic 2 Pro drone. Due to local flight regulations, we limited the drone flight altitude to 50 meters.
The first step was to capture the GPS locations and their corresponding video frames from the drone's video stream. To do so, we built an app that collects this information using DJI's Android mobile software development kit (SDK) \cite{DJI2020SDK}. The ability to downsample or oversample the video stream allows us to simulate a flight speed that differs from the actual speed of the drone. We gatheedr the samples, which will serve as input to our model, and stored them on a PC.

\subsection{The Experiment}
We performed drone flights in two urban areas (see the flight route for location \#1 and \#2 at an altitude of 50 meters in Figure \ref{fig:hod_ramat_GPS_50m} and at an altitude of 100 meters in Figure \ref{fig:hod_ramat_GPS_100m}). We performed two types of experiments; in one experiment the route was in the shape of a star (simulating a pizza delivery drone that returns to its base station after each delivery to collect a hot pizza), and in the other experiment a different, non-star route (simulating a case in which a few packages were given to the drone as might be done with an Amazon delivery drone) was used. We obtained a total of 5,496 frames at an altitude of 50 meters and 5,425 frames at an altitude of 100 meters, each of which was associated with a specific GPS location at a point in time. For simplicity, the drone flew at a speed of approximately 4 km/h, and one frame and its GPS location was obtained per second. 

A test window is defined as the number of points needed to construct a linear regression function so that $f(correlation) = distance\_prediction$. The subsequent frames can be predicted by correlation to the distance regarding the same window. In each experiment, we used different window sizes to create the linear regression and calculated the predicted difference in the GPS location distance. We focused on three future points and estimated their position based on the linear regression function, as doing so would allow us to raise an alert within a reasonable amount of time in the case of a GPS spoofing attack. Every calculation window utilizes resources, so we aimed to define a calculation window that optimizes the number of calculations and the false positive rate (FPR), which we would like to keep low.

\begin{figure}[t] 
\begin{center}
    \fbox{\includegraphics[width=.4\textwidth]{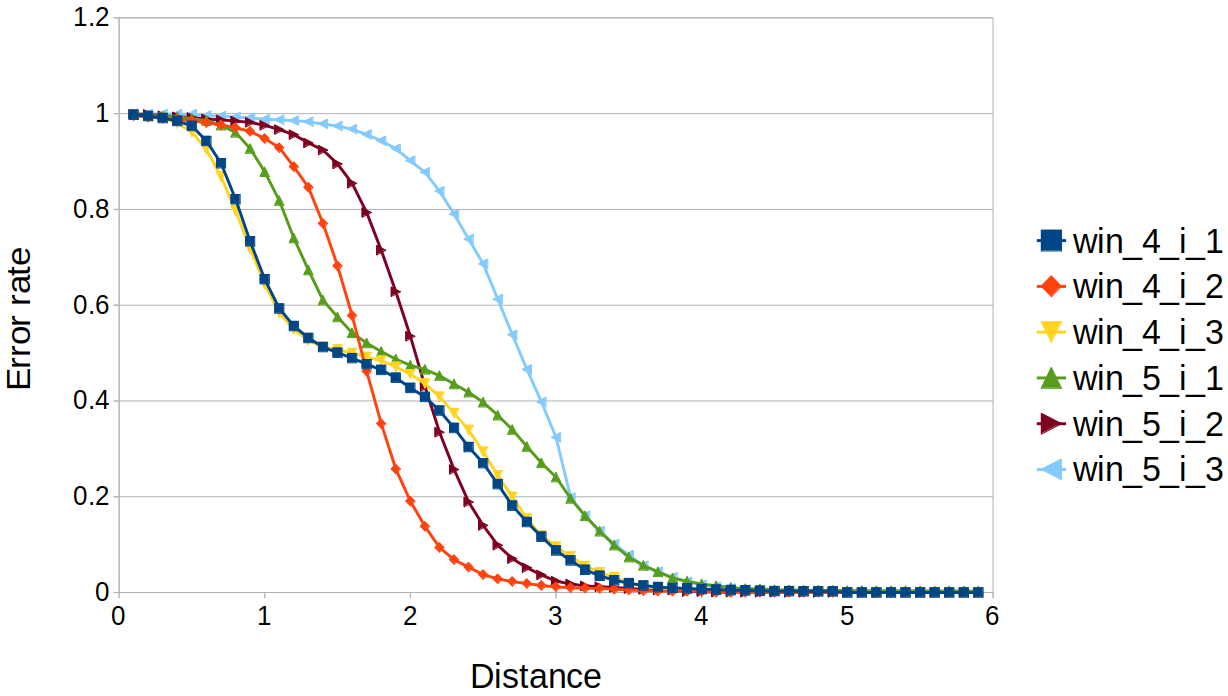}} \hspace{1PX}
    \caption{The FPR for a window size of 4 and 5 frames and an altitude of 50 meters.}
    \label{fig:FPR 50m}
\end{center}
\end{figure}

\begin{figure}[t] 
\begin{center}
    \fbox{\includegraphics[width=.4\textwidth]{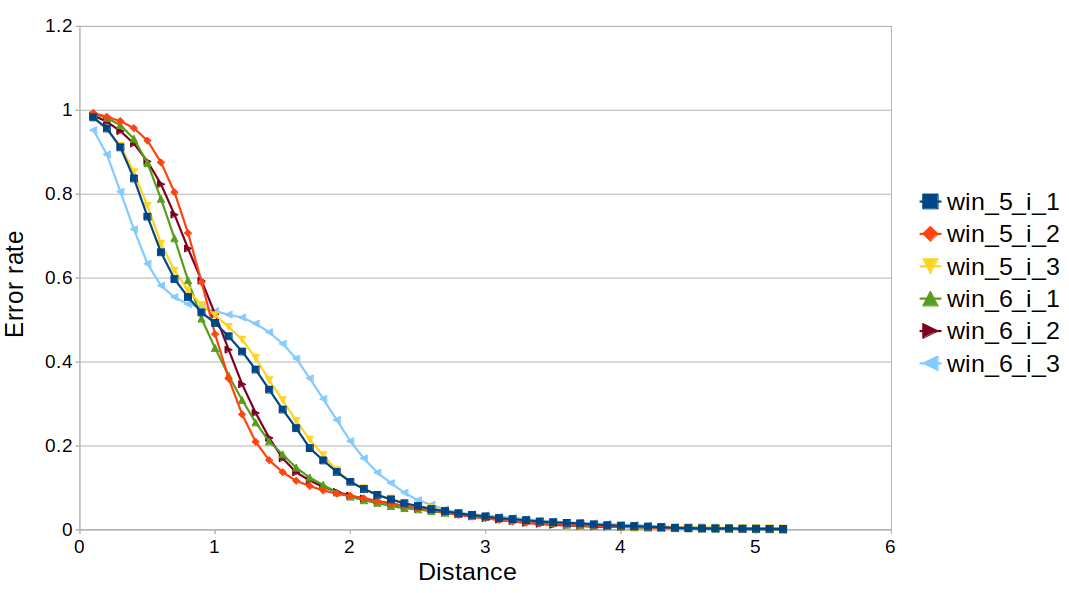}} \hspace{1PX}
    \caption{The FPR for a window size of 5 and 6 frames and an altitude of 100 meters.}
    \label{fig:FPR 100m}
\end{center}
\end{figure}

\begin{figure}[t] 
\begin{center}
    \fbox{\includegraphics[width=.4\textwidth]{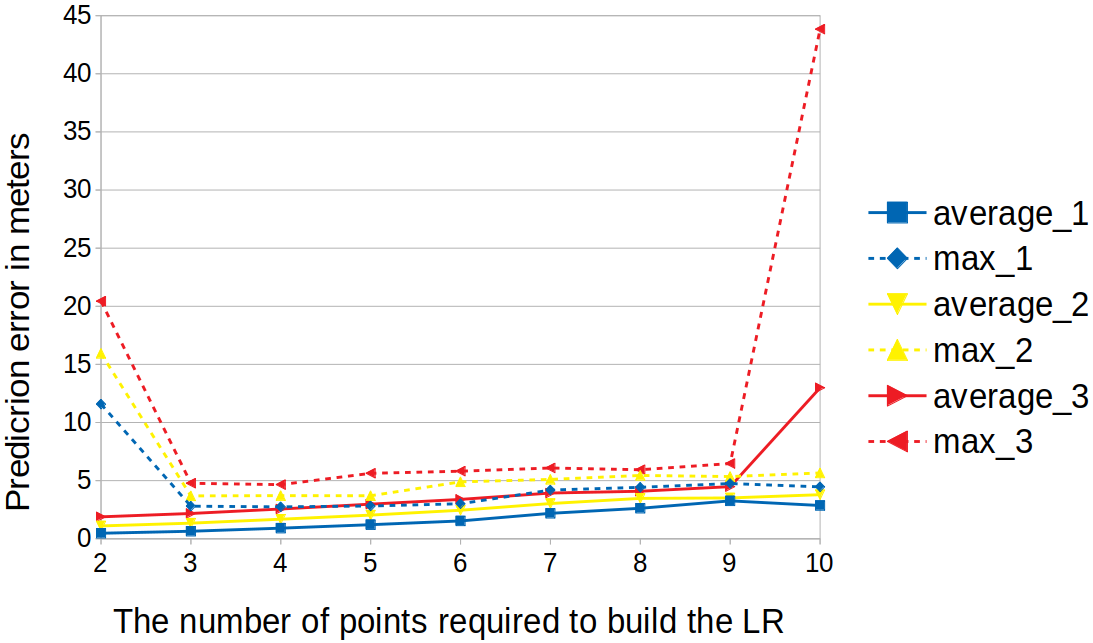}} \hspace{1PX}
    \caption{The prediction error (in meters) vs various window sizes and an altitude of 50 meters (average for the 2 locations).}
    \label{fig:Eval 50m hod and ramat}
\end{center}
\end{figure}

\begin{figure}[t] 
\begin{center}
    \fbox{\includegraphics[width=.4\textwidth]{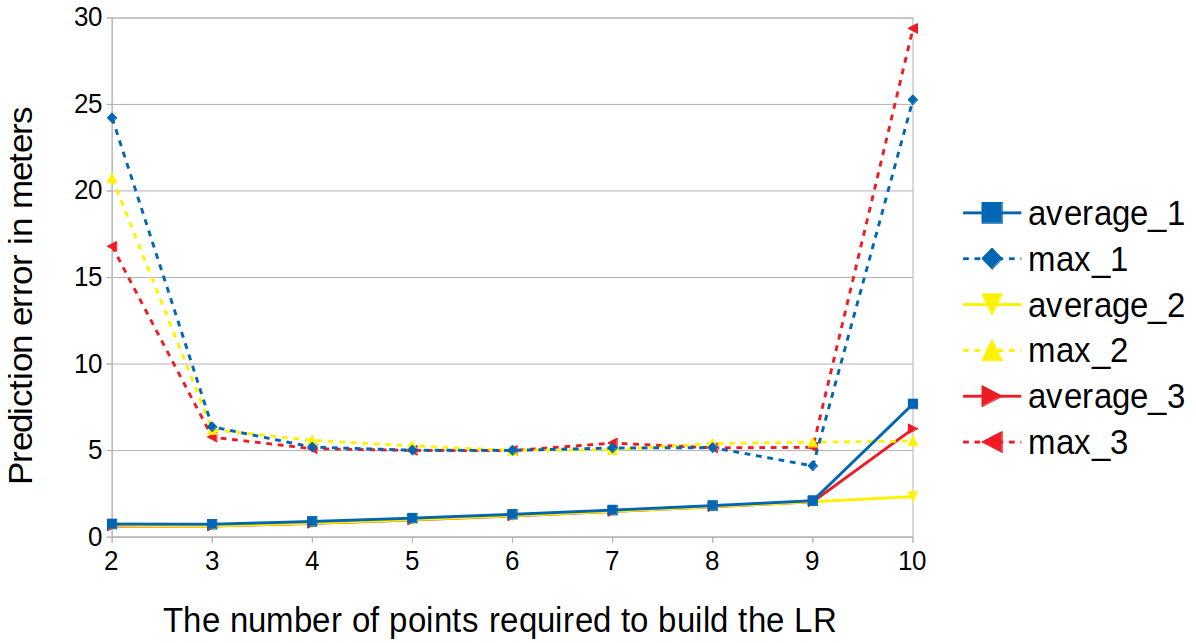}} \hspace{1PX}
    \caption{The prediction error (in meters) vs various window sizes and an altitude of 100 meters (average for the 2 locations).}
    \label{fig:Eval 100m hod and ramat}
\end{center}
\end{figure}

\subsection{Results}
We present the results of our experiments for a drone flying at an altitude of 50 and 100 meters. First, we calculate the average prediction error and maximum prediction error for each time window. Then, for each flight we plot the mean of the average prediction errors and maximum prediction errors at both altitudes (see Figures \ref{fig:Eval 50m hod and ramat} and \ref{fig:Eval 100m hod and ramat}). 

We observe that when the time window is small, the average prediction error is high. For example, in Figure \ref{fig:Eval 50m hod and ramat} the window size is two (meaning two frames are used to build the linear regression function); in this case, the average prediction error is less than one, but the maximum error is high. In addition, in the figures we can see that in the middle frames the maximum error is at the lowest level, and it remains constant for several window sizes; when the window size grows, the error also starts to increase.

Next, we examine the FPR for specific window sizes in which the maximum prediction error remains constant. At an altitude of 50 meters, the maximum error distance is six meters for all window sizes, and at an altitude of 100 meters, the maximum error distance is five meters for all window sizes. The best results at an altitude of 50 meters are for window sizes 4 and 5, where we obtain  the lowest error rate at a distance of four meters (Figures \ref{fig:FPR 50m}). While at 100 meters, the best results are for window sizes 5 and 6, where we obtain  the lowest error rate at a distance of 3.5 meters (see Figure \ref{fig:FPR 100m}).

We also investigate how we can determine the optimal window size for our method -- the size that will minimize the maximum error and average error. For that, we can use the equation: $e(\alpha) = \alpha*average_{error} + (1-\alpha)*maximum_{error}$. In Figures \ref{fig:Eval 50m hod and ramat} and \ref{fig:Eval 100m hod and ramat}, we can see that at an altitude of 50 meters the maximum error and average error are at window sizes 4 and 5, while at 100 meters they are at window sizes 5 and 6.

In conclusion, for the following configuration, the proposed method can provide a level of security that detects any GPS spoofing attack in which the spoofed location is a distance of 1-4 meters (an average of 2.5 meters) from the real location. Given this, we conclude that the proposed method is capable of protecting a delivery drone from GPS spoofing attacks, the method can be used for this operation.

\section{Summary \& Discussion}
\label{sec:summary}

In the next few years, drone use for commercial will increase. Given the ease with which GPS spoofing attacks can be performed, amateurs may try to attack delivery drones in order to steal the goods they transport. In this paper, we presented a method for the detection of GPS spoofing attacks using a drone's video stream. Our evaluation results demonstrate our method's ability protect delivery drones from GPS spoofing attacks. Our method's advantages include the fact that it does not require any extra hardware or prior knowledge on the flight area.

\bibliographystyle{IEEEtranS}

\bibliography{main}

\begin{thebibliography}{10}
\providecommand{\url}[1]{#1}
\csname url@samestyle\endcsname
\providecommand{\newblock}{\relax}
\providecommand{\bibinfo}[2]{#2}
\providecommand{\BIBentrySTDinterwordspacing}{\spaceskip=0pt\relax}
\providecommand{\BIBentryALTinterwordstretchfactor}{4}
\providecommand{\BIBentryALTinterwordspacing}{\spaceskip=\fontdimen2\font plus
\BIBentryALTinterwordstretchfactor\fontdimen3\font minus
  \fontdimen4\font\relax}
\providecommand{\BIBforeignlanguage}[2]{{%
\expandafter\ifx\csname l@#1\endcsname\relax
\typeout{** WARNING: IEEEtranS.bst: No hyphenation pattern has been}%
\typeout{** loaded for the language `#1'. Using the pattern for}%
\typeout{** the default language instead.}%
\else
\language=\csname l@#1\endcsname
\fi
#2}}
\providecommand{\BIBdecl}{\relax}
\BIBdecl

\bibitem{abera2019diat}
T.~Abera, R.~Bahmani, F.~Brasser, A.~Ibrahim, A.-R. Sadeghi, and M.~Schunter,
  ``Diat: Data integrity attestation for resilient collaboration of autonomous
  systems.'' in \emph{NDSS}, 2019.

\bibitem{DJI2020SDK}
\BIBentryALTinterwordspacing
DJI, ``Dji sdk platform,'' 2020. [Online]. Available:
  \url{https://developer.dji.com/mobile-sdk/}
\BIBentrySTDinterwordspacing

\bibitem{feng2018efficient}
Z.~Feng, N.~Guan, M.~Lv, W.~Liu, Q.~Deng, X.~Liu, and W.~Yi, ``An efficient uav
  hijacking detection method using onboard inertial measurement unit,''
  \emph{ACM Transactions on Embedded Computing Systems (TECS)}, vol.~17, no.~6,
  pp. 1--19, 2018.

\bibitem{jansen2016multi}
K.~Jansen, N.~O. Tippenhauer, and C.~P{\"o}pper, ``Multi-receiver gps spoofing
  detection: Error models and realization,'' in \emph{Proceedings of the 32nd
  Annual Conference on Computer Security Applications}, 2016, pp. 237--250.

\bibitem{kerns2014unmanned}
A.~J. Kerns, D.~P. Shepard, J.~A. Bhatti, and T.~E. Humphreys, ``Unmanned
  aircraft capture and control via gps spoofing,'' \emph{Journal of Field
  Robotics}, vol.~31, no.~4, pp. 617--636, 2014.

\bibitem{luo2016drones}
A.~Luo, ``Drones hijacking,'' \emph{DEF CON, Paris, France, Tech. Rep}, 2016.

\bibitem{nassi2019sok}
B.~Nassi, R.~Bitton, R.~Masuoka, A.~Shabtai, and Y.~Elovici, ``Sok: Security
  and privacy in the age of commercial drones,'' in \emph{Proc. IEEE Symp.
  Security Privacy (SP)}, 2021, pp. 73--90.

\bibitem{qiao2017vision}
Y.~Qiao, Y.~Zhang, and X.~Du, ``A vision-based gps-spoofing detection method
  for small uavs,'' in \emph{2017 13th International Conference on
  Computational Intelligence and Security (CIS)}.\hskip 1em plus 0.5em minus
  0.4em\relax IEEE, 2017, pp. 312--316.

\bibitem{ranganathan2016spree}
A.~Ranganathan, H.~{\'O}lafsd{\'o}ttir, and S.~Capkun, ``Spree: A spoofing
  resistant gps receiver,'' in \emph{Proceedings of the 22nd Annual
  International Conference on Mobile Computing and Networking}, 2016, pp.
  348--360.

\bibitem{wiki:Technology_readiness_level}
Wikipedia, ``{Technology readiness level} --- {W}ikipedia{,} the free
  encyclopedia,''
  \url{http://en.wikipedia.org/w/index.php?title=Technology\%20readiness\%20level&oldid=1060145453},
  2021, [Online; accessed 28-December-2021].

\bibitem{xue2020deepsim}
N.~Xue, L.~Niu, X.~Hong, Z.~Li, L.~Hoffaeller, and C.~P{\"o}pper, ``Deepsim:
  Gps spoofing detection on uavs using satellite imagery matching,'' in
  \emph{Annual Computer Security Applications Conference}, 2020, pp. 304--319.

\end{thebibliography}

\end{document}